\documentclass[twocolumn]{aastex62}
\usepackage{natbib}
\usepackage{epsfig}
\usepackage{graphicx}
\usepackage{subfigure}
\usepackage{float}
\usepackage{amsmath}
\usepackage{color}
\usepackage{amssymb}
\usepackage{amsfonts}
\usepackage{apjfonts}

\newcommand{\PMO}{Purple Mountain Observatory, Chinese Academy of Sciences, Nanjing 210023, China}
\newcommand{\GXU}{Guangxi Key Laboratory for Relativistic Astrophysics, Nanning 530004, China}
\newcommand{\USTC}{School of Astronomy and Space Sciences, University of Science and Technology of China, Hefei 230026, China}
\newcommand{\NAOC}{Key Laboratory for Computational Astrophysics, National Astronomical Observatories, Chinese Academy of Sciences, Beijing 100101, China}
\newcommand{\BNU}{Department of Astronomy, Beijing Normal University, Beijing 100875, China}

\shortauthors{Wei et al.}

\begin{document}

\title{Direct Estimate of the Post-Newtonian Parameter and Cosmic Curvature from Galaxy-scale Strong Gravitational Lensing}

\correspondingauthor{Jun-Jie Wei, Xue-Feng Wu}
\email{jjwei@pmo.ac.cn, xfwu@pmo.ac.cn}

\author{Jun-Jie Wei}
\affiliation{\PMO}
\affiliation{\GXU}
\affiliation{\USTC}

\author{Yun Chen}
\affiliation{\NAOC}

\author{Shuo Cao}
\affiliation{\BNU}

\author{Xue-Feng Wu}
\affiliation{\PMO}
\affiliation{\USTC}

\begin{abstract}

Einstein's theory of general relativity (GR) has been precisely tested on solar system scales,
but extragalactic tests are still poorly performed. In this work, we use a newly compiled sample
of galaxy-scale strong gravitational lenses to test the validity of GR on kiloparsec scales.
In order to solve the circularity problem caused by the preassumption of a specific cosmological
model based on GR, we employ the distance sum rule in the Friedmann-Lema\^{\i}tre-Robertson-Walker
metric to directly estimate the parameterized post-Newtonian (PPN) parameter $\gamma_{\rm PPN}$
and the cosmic curvature $\Omega_k$ by combining observations of strong lensing and Type Ia
supernovae. This is the first simultaneous measurement of $\gamma_{\rm PPN}$ and $\Omega_k$ without
any assumptions about the contents of the universe or the theory of gravity. Our results show that
$\gamma_{\rm PPN}=1.11^{+0.11}_{-0.09}$ and $\Omega_{k}=0.48^{+1.09}_{-0.71}$, indicating a strong
degeneracy between the two quantities. The measured $\gamma_{\rm PPN}$, which is consistent with
the prediction of 1 from GR, provides a precise extragalactic test of GR with a fractional accuracy
better than 9.0\%. If a prior of the spatial flatness (i.e., $\Omega_{k}=0$) is adopted, the PPN parameter
constraint can be further improved to $\gamma_{\rm PPN}=1.07^{+0.07}_{-0.07}$, representing a
precision of 6.5\%. On the other hand, in the framework of GR (i.e., $\gamma_{\rm PPN}=1$),
our results are still marginally compatible with zero curvature ($\Omega_k=-0.12^{+0.48}_{-0.36}$),
supporting no significant deviation from a flat universe.

\end{abstract}


\section{Introduction}

Einstein's theory of general relativity (GR) is one of the major pillars of modern physics. Any possible violation of GR
would have far-reaching consequences for our understanding of fundamental physics; testing GR at a much higher precision
has therefore been one of the most enduring pursuits of scientists. At the post-Newtonian level, the validity of GR can be
tested by constraining the parameterized post-Newtonian (PPN) parameter $\gamma_{\rm PPN}$, since GR predicts exactly
$\gamma_{\rm PPN}\equiv1$ \citep{1971ApJ...163..595T,2006LRR.....9....3W,2014LRR....17....4W}. Here, $\gamma_{\rm PPN}$
stands for the amount of space-curvature generated by a unit rest mass.
On solar system scales, tests of GR through numerical values of $\gamma_{\rm PPN}$ have reached high precision.
By measuring the arrival-time delay of radar signals passing close to the Sun, the Cassini spacecraft yielded an agreement
with GR to $10^{-3}\%$, i.e., $\gamma_{\rm PPN}=1+(2.1\pm2.3)\times10^{-5}$ \citep{2003Natur.425..374B}. However,
current extragalactic tests of GR are much less precise. On scales of 10--100 Mpc, only $\sim20$\% precision
on the constraints of $\gamma_{\rm PPN}$ has been obtained using the joint measurements of weak gravitational lensing
and redshift-space distortions \citep{2011PhRvD..84h3523S,2013MNRAS.429.2249S,2016MNRAS.456.2806B}. On megaparsec scales,
$\gamma_{\rm PPN}$ has been limited to just 30\% precision by analyzing the mass profiles of galaxy clusters
\citep{2015MNRAS.452.1171W,2016JCAP...04..023P}.

On kiloparsec scales, strong gravitational lensing (SGL) systems, combined with stellar dynamical data of lensing galaxies,
provide an effective tool to verify the weak-field metric of gravity. For a specific SGL system with the foreground galaxy
acting as a lens, multiple images, arcs, or even an Einstein ring can form with angular separations close to the so-called
Einstein radius \citep{2017JCAP...07..045C}.
In theory, the Einstein radius is related to the mass of the lens, the PPN parameter $\gamma_{\rm PPN}$,
and a ratio of three angular diameter distances (i.e., the distances from the observer to the lens and the source, $D_l$
and $D_s$, and the distance between the lens and the source $D_{ls}$) \citep{Cao_2015}. With the required angular diameter
distances and measurements of the lens mass and the Einstein radius, one can therefore constrain $\gamma_{\rm PPN}$ and
test whether GR is a suitable theory of gravity on the corresponding scales.
This method was first performed on 15 lensing galaxies from the Sloan Lens ACS Survey by \cite{2006PhRvD..74f1501B},
which yielded $\gamma_{\rm PPN}=0.98\pm0.07$ based on prior assumptions on galaxy structure from local observations.
Subsequently, different SGL samples have been used to test the accuracy of GR \citep{2009arXiv0907.4829S,2010ApJ...708..750S,
2017ApJ...835...92C,2018Sci...360.1342C,2020MNRAS.497L..56Y,2021arXiv210902291L}. In most previous studies, the distance
information required to constrain the PPN parameter $\gamma_{\rm PPN}$ is provided by the prediction of the standard
$\Lambda$CDM cosmological model. It should, however, be emphasized that $\Lambda$CDM is established based on the framework
of GR. Thus, there is a circularity problem in testing GR \citep{2021arXiv210902291L}. To overcome this problem, one has to
determine the lensing distance ratio in a cosmology-independent way.

The circularity problem can be alleviated by determining the two distances $D_l$ and $D_s$ through observations of Type Ia
supernovae (SNe Ia). But, the distance $D_{ls}$ cannot be determined directly from the observations. In the
Friedmann-Lema\^{\i}tre-Robertson-Walker (FLRW) metric, these three distances are related via the distance sum rule (DSR),
which depends on the curvature parameter of the universe $\Omega_{k}$. Turning this around, supposing that
the universe is described by the FLRW metric, we can use combined observations of strong lensing and SNe Ia to estimate
not only $\gamma_{\rm PPN}$ but also $\Omega_{k}$ independently of the cosmological model \citep{2017ApJ...835...92C}.
Based on the DSR in the FLRW metric, and assuming that GR is valid (i.e., $\gamma_{\rm PPN}=1$),
model-independent constraints on the cosmic curvature $\Omega_{k}$ have been implemented by combining SGL systems
with other distance indicators \citep{2015PhRvL.115j1301R,2017ApJ...839...70L,2017ApJ...834...75X,2018JCAP...03..041D,
2018ApJ...854..146L,2018NatCo...9.3833L,2019ApJ...873...37L,2019NatSR...911608C,2021arXiv211200237C,2019PhRvL.123w1101C,
2019PhRvD..99h3514L,2019PhRvD.100b3530Q,2020MNRAS.496..708L,2019MNRAS.483.1104Q,2021MNRAS.503.2179Q,2020ApJ...898..100W,2020ApJ...897..127W,
2020ApJ...889..186Z,2021MNRAS.506L...1D}. Without the prior assumption on GR, \cite{2017ApJ...835...92C} proposed that this
cosmology-independent method could be extended to study the degeneracy between the PPN parameter $\gamma_{\rm PPN}$
and the curvature parameter $\Omega_{k}$. They used the simulated strong-lensing data to estimate both $\gamma_{\rm PPN}$
and $\Omega_{k}$. We will now for the first time apply such a method to real data.

We should note that a recent work by \cite{2021arXiv210902291L} used strong lensing and SNe Ia to obtain model-independent
constraints on $\gamma_{\rm PPN}$ within the framework of the flat FLRW metric (i.e., $\Omega_{k}=0$). However,
\cite{2017ApJ...835...92C} proved that there exists a significant degeneracy between $\gamma_{\rm PPN}$ and $\Omega_{k}$
by simulation. Obviously, a simple flatness assumption may lead to a biased estimate of $\gamma_{\rm PPN}$, even if
the real curvature is tiny. Therefore, it would be better to simultaneously optimize $\gamma_{\rm PPN}$ and $\Omega_{k}$,
as we do in this work.

The outline of this work is as follows. In Section~\ref{sec:method}, we introduce the gravitational lensing theory and
the DSR method. In Section~\ref{sec:data}, we describe the observational data used for our analysis.
Model-independent constraints on $\gamma_{\rm PPN}$ and $\Omega_{k}$ are presented in Section~\ref{sec:constraint}.
Finally, a brief summary and discussions are given in Section~\ref{sec:summary}.

\section{Methodology}
\label{sec:method}
In the limit of a weak gravitational field, the general form of the Schwarzschild metric for a point mass
$M$ can be written as
\begin{equation}
\mathrm{d}s^{2}=c^{2}\mathrm{d}t^{2}\left(1-\frac{2GM}{c^{2}r}\right)-\mathrm{d}r^{2}\left(1+\frac{2\gamma_{\rm PPN} GM}{c^{2}r}\right)-r^{2}\mathrm{d}\Omega^{2}\;,
\end{equation}
where $\gamma_{\rm PPN}$ is the PPN parameter and $\Omega$ is the angle in the invariant orbital plane.
In GR, $\gamma_{\rm PPN}$ is predicted to be $1$.

\subsection{Gravitational Lensing Theory}
\label{subsec:SGL}
The core idea of using the SGL systems to test gravity is that the gravitational mass $M_{\rm E}^{\rm grl}$ and
the dynamical mass $M_{\rm E}^{\rm dyn}$ enclosed within the Einstein ring should be equivalent, i.e.,
\begin{equation}\label{eq:equalM}
M_{\rm E}^{\rm grl}=M_{\rm E}^{\rm dyn}\;.
\end{equation}
From the theory of gravitational lensing, the gravitational mass $M_{\rm E}^{\rm grl}$ is related to
the Einstein angle $\theta_{\rm E}$ (reflecting the angular separation between multiple images; \citealt{2017ApJ...835...92C})
\begin{equation}
\theta_{\rm E}=\sqrt{\frac{1+\gamma_{\rm PPN}}{2}}\left(\frac{4G M_{\rm E}^{\rm grl}}{c^2}\frac{D_{ls}}{D_{s}D_{l}}\right)^{1/2}\;,
\end{equation}
where $D_{s}$ is the angular diameter distance to the source, $D_{l}$ is the angular diameter distance to the lens,
and $D_{ls}$ is the angular diameter distance between the lens and the source \citep{Cao_2015}. By substituting the
Einstein ring radius $R_{\rm E}=\theta_{\rm E}D_{l}$, one can further figure out
\begin{equation}\label{eq:Mgrl}
\frac{G M_{\rm E}^{\rm grl}}{R_{\rm E}}=\frac{2}{\left(1+\gamma_{\rm PPN}\right)}\frac{c^{2}}{4}\frac{D_{s}}{D_{ls}}\theta_{\rm E}\;.
\end{equation}

Given the mass distribution model for the lensing galaxy, the dynamical mass $M_{\rm E}^{\rm dyn}$ can be inferred
from the spectroscopic measurement of the lens velocity dispersion. Here we adopt a general mass model with
power-law density profiles for the lensing galaxy \citep{2006EAS....20..161K,2016MNRAS.461.2192C}:
\begin{equation}\label{eq:lens}
\left\{\begin{array}{l}
\rho(r)=\rho_{0}\left(r / r_{0}\right)^{-\alpha} \\
\nu(r)=\nu_{0}\left(r / r_{0}\right)^{-\delta} \\
\beta(r)=1-\sigma_{t}^{2} / \sigma_{r}^{2}\;,
\end{array}\right.
\end{equation}
where $r$ is the spherical radial coordinate from the lens center, $\rho(r)$ is the total (i.e., luminous plus
dark matter) mass density, and $\nu(r)$ denotes the luminosity density of stars. The parameter $\beta(r)$
represents the anisotropy of the stellar velocity dispersion, which relates to the velocity dispersions, $\sigma^{2}_{t}$
and $\sigma^{2}_{r}$, in the tangential and radial directions. Also, $\alpha$ and $\delta$ are the slopes of the
power-law density profiles. It is worth noting that the total mass density slope $\alpha$ is significantly dependent
on both the lens redshift $z_l$ and the surface mass density (e.g., \citealt{2013ApJ...777...98S,2019MNRAS.488.3745C}).
\cite{2019MNRAS.488.3745C} proved that the most compatible lens mass model is
\begin{equation}
\alpha=\alpha_{0}+\alpha_{z}z_{l}+\alpha_{s}\log_{10}\tilde{\Sigma}\;,
\end{equation}
where $\alpha_{0}$, $\alpha_{z}$, and $\alpha_{s}$ are free parameters.
Here $\tilde{\Sigma}$ denotes the normalized surface mass density of the lensing galaxy, which is given by
$\tilde{\Sigma}=\frac{\left(\sigma_{0} / 100 \rm{~km} \rm{~s}^{-1}\right)^{2}}{R_{\rm{eff}} / 10~h^{-1} \rm{~kpc}}$,
where $\sigma_0$ is the observed velocity dispersion, $h=H_0/(100 \rm{~km} \rm{~s}^{-1} \rm{~Mpc}^{-1})$
is the reduced Hubble constant, and $R_{\mathrm{eff}}$ is the half-light radius of the lensing galaxy.
In the literature, the velocity anisotropy parameter $\beta$ is usually assumed to be independent of $r$
(e.g., \citealt{2006ApJ...649..599K,2010ApJ...709.1195T}). From a well-studied sample of nearby elliptical galaxies
\citep{2001AJ....121.1936G}, the posterior probability of $\beta$ is found to be characterized by a Gaussian
distribution, $\beta=0.18\pm0.13$, that is extensively adopted in previous works (e.g.,
\citealt{2006PhRvD..74f1501B,2010ApJ...708..750S,2017ApJ...835...92C,2019MNRAS.488.3745C,2021arXiv210902291L}).
Following these previous works, we will marginalize the anisotropy parameter $\beta$ using a Gaussian prior of
$\beta=0.18\pm0.13$ over the range of $\left[\bar{\beta}-2 \sigma_{\beta}, \bar{\beta}+2 \sigma_{\beta}\right]$,
where $\bar{\beta}=0.18$ and $\sigma_{\beta}=0.13$.

Based on the radial Jeans equation in spherical coordinate, the radial velocity dispersion of luminous matter
in early-type lens galaxies can be expressed as
\begin{equation} \label{eq:sigma}
\sigma_{r}^{2}(r)=\frac{G \int_{r}^{\infty} \mathrm{d} r^{\prime} r^{\prime 2 \beta-2} \nu(r^{\prime}) M(r^{\prime})}{r^{2 \beta} \nu(r)}\;,
\end{equation}
where $M(r)$ is the total mass contained within a spherical radius $r$.
With the mass density profiles in Equation~(\ref{eq:lens}), we can derive the
relation between the dynamical mass $M_{\rm E}^{\rm dyn}$ enclosed within the Einstein ring radius $R_{\rm E}$ and $M(r)$ as
(see \citealt{2006EAS....20..161K,2019MNRAS.488.3745C} for the detailed derivation)
\begin{equation} \label{eq:Mr}
M(r)=\frac{2}{\sqrt{\pi}} \frac{1}{\lambda(\alpha)} \left(\frac{r}{R_{\rm E}}\right)^{3-\alpha} M_{\rm E}^{\rm dyn}\;,
\end{equation}
where $\lambda(x)=\Gamma\left(\frac{x-1}{2}\right)/\Gamma\left(\frac{x}{2}\right)$ stands for the ratio of two respective
Gamma functions. By substituting Equations~(\ref{eq:Mr}) and (\ref{eq:lens}) into Equation~(\ref{eq:sigma}), one can have
\begin{equation} \label{sigma_r}
\sigma_{r}^{2}(r)=\frac{2}{\sqrt{\pi}} \frac{G M_{\rm E}^{\rm dyn}}{R_{\rm E}} \frac{1}{\xi-2 \beta} \frac{1}{\lambda(\alpha)}\left(\frac{r}{R_{\rm E}}\right)^{2-\alpha}\;,
\end{equation}
where $\xi=\alpha+\delta-2$.

The actual velocity dispersion of the lensing galaxy is effectively averaged by line-of-sight luminosity and measured over
the effective spectroscopic aperture $R_{\rm A}$, which can be expressed as (see \citealt{2019MNRAS.488.3745C} for the detailed
derivation)
\begin{equation}\label{eq:sigma_RA1}
\sigma^{2}_{0}(\leq R_{\rm A})=\frac{2}{\sqrt{\pi}} \frac{G M_{\rm E}^{\rm dyn}}{R_{\rm E}} F(\alpha, \delta, \beta) \left(\frac{R_{\rm A}}{R_{\rm E}}\right)^{2-\alpha}\;,
\end{equation}
where
\begin{equation}
F(\alpha, \delta, \beta)= \frac{3-\delta}{\left(\xi-2 \beta\right)\left(3-\xi\right)}\frac{\lambda(\xi)-\beta\lambda(\xi+2)}{\lambda(\alpha)\lambda(\delta)} \;.
\end{equation}
Lastly, with the relations expressed in Equations~(\ref{eq:equalM}) and (\ref{eq:Mgrl}), Equation~(\ref{eq:sigma_RA1})
can be rewritten as
\begin{equation}\label{eq:sigma_RA2}
\sigma^{2}_{0}(\leq R_{\rm A})=\frac{c^{2}}{2\sqrt{\pi}} \frac{2}{\left(1+\gamma_{\rm PPN}\right)} \frac{D_{s}}{D_{ls}}\theta_{\rm E} F(\alpha, \delta, \beta) \left(\frac{\theta_{\rm A}}{\theta_{\rm E}}\right)^{2-\alpha}\;,
\end{equation}
where $R_{\rm A}=\theta_{\rm A}D_{l}$.

From the spectroscopic data, one can measure the lens velocity dispersion $\sigma_{\rm ap}$ inside the circular
aperture with the angular radius $\theta_{\rm ap}$. In practice, the luminosity-weighted
average of the line-of-sight velocity dispersion $\sigma_{\rm ap}$ measured within a certain aperture should be normalized
to a typical physical aperture with the radius $\theta_{\rm eff}/2$,
\begin{equation} \label{eq:sigma_obs}
\sigma^{\rm obs}_{0}=\sigma_{\mathrm{ap}}\left[\theta_{\mathrm{eff}} /\left(2 \theta_{\mathrm{ap}}\right)\right]^{\eta}\;,
\end{equation}
where $\theta_{\rm eff}=R_{\rm eff}/D_{l}$ is the effective angular radius of the lensing galaxy. Following
\cite{2019MNRAS.488.3745C}, we adopt the value of the correction factor $\eta=-0.066\pm0.035$ from \cite{2006MNRAS.366.1126C}.
Then, we can calculate the total uncertainty of $\sigma^{\rm obs}_{0}$ using the expression
\begin{equation}\label{eq:sigma_error}
\left(\Delta \sigma_{0}^{\mathrm{tot}}\right)^{2}=\left(\Delta \sigma_{0}^{\mathrm{stat}}\right)^{2}+\left(\Delta \sigma_{0}^{\mathrm{AC}}\right)^{2}+\left(\Delta \sigma_{0}^{\mathrm{sys}}\right)^{2}\;,
\end{equation}
where $\Delta \sigma_{0}^{\mathrm{stat}}$ is the statistical uncertainty propagated from the measurement error of $\sigma_{\rm{ap}}$.
The uncertainty caused by the aperture correction, $\Delta \sigma_{0}^{\mathrm{AC}}$, is propagated from
the error of $\eta$. The extra mass contribution from other matters (outside of the lensing galaxy) along the line of sight
in the estimation of $M_{\rm E}^{\rm grl}$ can be treated as a systematic uncertainty $\Delta \sigma_{0}^{\mathrm{sys}}$,
which contributes an uncertainty of $\sim3\%$ to the velocity dispersion \citep{2007ApJ...671.1568J}.

With Equation~(\ref{eq:sigma_RA2}), the theoretical value of the velocity dispersion within the radius $\theta_{\rm eff}/2$
takes the form \citep{2006EAS....20..161K}
\begin{equation}
\sigma^{\rm th}_{0}=\sqrt{\frac{c^{2}}{2\sqrt{\pi}} \frac{2}{\left(1+\gamma_{\rm PPN}\right)} \frac{D_{s}}{D_{ls}}\theta_{\rm E} F(\alpha, \delta, \beta) \left(\frac{\theta_{\rm eff}}{2\theta_{\rm E}}\right)^{2-\alpha}}\;.
\label{eq:sigma_theory}
\end{equation}
For the case of $\alpha=\delta=2$ and $\beta=0$, the mass model is reduced to the singular isothermal sphere (SIS) model,
and the theoretical value of the velocity dispersion is simplified as
$\sigma_{\rm SIS}=\sqrt{\frac{c^{2}}{4\pi}\frac{2}{\left(1+\gamma_{\rm PPN}\right)}\frac{D_{s}}{D_{ls}}\theta_{\rm E}}$.

By comparing the observational values of the velocity dispersions (Equation~(\ref{eq:sigma_obs})) with the corresponding
theoretical ones (Equation~(\ref{eq:sigma_theory})), one can place constraints on the PPN parameter $\gamma_{\rm PPN}$.
For this purpose, it is also necessary to know the distance ratio $D_{s}/D_{ls}$, which is conventionally calculated
in the context of flat $\Lambda$CDM \citep{2010ApJ...708..750S,2017ApJ...835...92C}. However, a circularity problem
exists in this approach because the standard $\Lambda$CDM cosmological model is built on the framework of GR
\citep{2021arXiv210902291L}. In order to avoid the circularity problem, we will apply a cosmology-independent method
to constrain $\gamma_{\rm PPN}$. This method is based on the sum rule of distances along null geodesics of the FLRW metric.

\subsection{Distance Sum Rule}
If space is exactly homogeneous and isotropic, the FLRW metric can be used to describe the spacetime geometry of the universe.
In the FLRW metric, the dimensionless comoving distance $d(z_l,z_s)\equiv (H_{0}/c)(1+z_s) D_A(z_l,z_s)$
is given by
\begin{equation}
d(z_l,z_s)=\frac{1}{\sqrt{|\Omega_{k}|}}{\rm sinn}\left(\sqrt{|\Omega_{k}|}\int_{z_l}^{z_s}\frac{{\rm d}z'}{E(z')}\right)\;,
\label{eq:dls}
\end{equation}
where $\Omega_{k}$ is the curvature parameter and $E(z)=H(z)/H_0$ is the dimensionless Hubble parameter.
Also, ${\rm sinn}(x)=\sinh(x)$ for $\Omega_{k}>0$ and ${\rm sinn}(x)=\sin(x)$ for $\Omega_{k}<0$. For a flat universe
with $\Omega_{k}=0$, Equation~(\ref{eq:dls}) reduces to a linear function of the integral. For an SGL system with the
notations $d(z)\equiv d(0,z)$, $d_l\equiv d(0,z_l)$, $d_s\equiv d(0,z_s)$, and $d_{ls}\equiv d(z_l,z_s)$, a simple
sum rule of distances in the FLRW framework can be easily derived as
\citep{1993ppc..book.....P,2006ApJ...637..598B,2015PhRvL.115j1301R}:
\begin{equation}
\frac{d_{ls}}{d_s}= \sqrt{1+\Omega_{k}d_{l}^{2}}-\frac{d_{l}}{d_s} \sqrt{1+\Omega_{k}d_{s}^{2}} \;.
\label{eq:DSR}
\end{equation}
This relation is very general because it only assumes that geometrical optics holds and that light propagation
is described with the FLRW metric. Once the derived $\Omega_{k}$ from the three distances ($d_l$, $d_s$, and $d_{ls}$)
is observationally found to be different for any two pairs of ($z_l$, $z_s$), we can rule out the FLRW metric.

Given independent measurements of $d_l$ and $d_s$ on the right side of Equation~(\ref{eq:DSR}), we are able to access
the dimensionless distance ratio $d_{ls}/d_{s}$,\footnote{Note that $d_{ls}/d_{s}$ is just equal to the ratio of
the angular diameter distances $D_{ls}/D_{s}$.} depending only on the curvature parameter $\Omega_{k}$ \citep{Geng_2020,2020MNRAS.496..708L,2021EPJC...81...14Z}.
Therefore, we can directly determine $\gamma_{\rm PPN}$ and $\Omega_{k}$ from Equations (\ref{eq:sigma_theory}) and
(\ref{eq:DSR}) without involving any specific cosmological model.

\section{Observational Data}
\label{sec:data}

\subsection{Supernova Data: The Distances $d_{l}$ and $d_s$}
In order to obtain model-independent estimate of $\gamma_{\rm PPN}$ and $\Omega_{k}$ via Equations (\ref{eq:sigma_theory})
and (\ref{eq:DSR}), we need to know the distances $d_l$ and $d_s$ on the right-hand-side terms of Equation~(\ref{eq:DSR}).
In principle, we can use different kinds of distance indicators such as standard candles, sirens, and rulers for
providing these two distances. Here, we use SN Ia observations to obtain $d_l$ and $d_s$.

\cite{2018ApJ...859..101S} released the largest combined sample of SNe Ia called Pantheon, which contains 1,048 SNe
in the redshift range $0.01<z<2.3$. Generally, the observed distance modulus of each SN is given by
$\mu_{\rm SN}=m_{B}+ \kappa \cdot X_{1}-\omega \cdot \mathcal{C}-M_{B}$, where $m_B$ is the observed peak magnitude
in the rest-frame \emph{B} band, $X_{1}$ and $\mathcal{C}$ are the light-curve stretch factor and the SN
color at maximum brightness, respectively, and $M_{B}$ is a nuisance parameter that represents the absolute
\emph{B}-band magnitude of a fiducial SN. Here, $\kappa$ and $\omega$ are two light-curve parameters, which could be
calibrated to zero through a method called BEAMS with Bias Corrections (BBC; \citealt{2017ApJ...836...56K}).
With the BBC method, \cite{2018ApJ...859..101S} reported the corrected apparent magnitudes
$m_{\rm corr}=\mu_{\rm SN}+M_{B}$ for all SNe. Therefore, the observed distance moduli $\mu_{\rm SN}$ can be
directly obtained by subtracting $M_{B}$ from $m_{\rm corr}$.

As proposed in \cite{2015PhRvL.115j1301R}, we determine the dimensionless distances $d_l$ and $d_s$ by fitting
a polynomial to the Pantheon SN Ia data. Here, we parameterize the dimensionless distance function as a third-order
polynomial with initial conditions $d(0)=0$ and $d'(0)=1$, i.e.,
\begin{equation}
d(z)=z+a_{1}z^2+a_{2}z^3\;,
\label{eq:dz}
\end{equation}
where $a_1$ and $a_2$ are two free parameters that need to be optimized along with the absolute magnitude $M_{B}$.
We find that higher-order polynomials do not improve the fitting performance, taking into account the larger number
of free parameters. That is, a simple third-order polynomial is flexible enough to fit the SN Ia data.

Given a vector of distance residuals of the Pantheon SN sample that may be expressed
as $\Delta \bf{\hat{\mu}}=\bf{\hat{\mu}_{\rm SN}}-\bf{\hat{\mu}_{\rm model}}$,
where $\bf{\hat{\mu}_{\rm SN}}$ ($\bf{\hat{\mu}_{\rm model}}$) is the observed (model) vector of distance moduli,
the likelihood for the model fit is defined by
\begin{equation}
-2 \ln\left(\mathcal{L}_{\rm SN}\right) = \Delta \bf{\hat{\mu}}^{\emph{T}} \cdot \textbf{Cov}^{-1} \cdot \Delta \bf{\hat{\mu}}\;,
\end{equation}
where $\textbf{Cov}$ is a covariance matrix that includes both statistical and systematic uncertainties of SNe.
Here the observed vector $\bf{\hat{\mu}_{\rm SN}}$ is given by $\mu_{{\rm SN},i}=m_{{\rm corr},i}-M_{B}$,
and the model vector $\bf{\hat{\mu}_{\rm model}}$ is determined by
$\mu_{{\rm model},i}=5\log_{10}[D_{L}(z_i)/{\rm 10\; pc}]=5\log_{10}[(1+z_i)d(z_i)]-5\log_{10}({\rm 10\; pc}\;H_0/c)$.
Given the degeneracy between the absolute magnitude $M_{B}$ and the Hubble constant $H_0$, we adopt a fiducial
$H_0=70$ km $\rm s^{-1}$ $\rm Mpc^{-1}$ for the sake of optimizing $M_{B}$.

\subsection{Strong-lensing Data: The Distance Ratio $d_{ls}/d_{s}$}
According to the analysis in Section~\ref{subsec:SGL}, one can learn that the underlying method requires the following
observational information of each SGL system, including the source redshift $z_s$, the lens redshift $z_l$, the Einstein
angle $\theta_{\rm E}$, the half-light angular radius of the lensing galaxy $\theta_{\rm eff}$, the spectroscopic aperture
angular radius $\theta_{\rm ap}$, and the lens velocity dispersion $\sigma_{\rm ap}$ measured within $\theta_{\rm ap}$.

Recently, \cite{2019MNRAS.488.3745C} compiled a sample of 161 galaxy-scale SGL systems with gravitational lensing
and stellar velocity dispersion measurements. In this sample, the slopes of the luminosity density profile $\delta$ of
130 SGL systems were measured by fitting the two-dimensional power-law luminosity profile convolved with the instrumental
point spread function to imaging data over a circle of radius $\theta_{\rm eff}/2$ centered on the lens galaxies.
By constraining the cosmological parameter $\Omega_{\rm m}$ separately with the entire sample of 161 SGL systems
(treating $\delta$ as a universal parameter for all lenses) and the truncated sample of 130 systems (treating
$\delta$ as an observable for each lens), \cite{2019MNRAS.488.3745C} suggested that the intrinsic scatter $\delta$ among
the lenses should be considered in order to get an unbiased estimate of $\Omega_{\rm m}$. Therefore, we adopt
this truncated sample of 130 SGL systems with $\delta$ measurements for the analysis demonstrated in this paper.
The redshift ranges of lens and source galaxies of these 130 SGL systems are $0.0624\leq z_{l}\leq0.7224$ and
$0.1970\leq z_{s}\leq2.8324$, respectively.

One of the limitations we must deal with in using the SGL data, however, is that the SN Ia measurements extend only to
$z=2.3$. As such, only a subset of the SGL sample that overlaps with the SN Ia catalog is actually available.
Our analysis will therefore be based only on the 120 SGL systems with $z_{s}<2.3$. The likelihood function
for strong-lensing data is then constructed as
\begin{equation}
\mathcal{L}_{\rm SGL}=\prod_{i=1}^{120} \frac{1}{\sqrt{2\pi}\,\Delta \sigma_{0,i}^{\mathrm{tot}}}
\exp\left[-\frac{1}{2}\left(\frac{\sigma^{\rm th}_{0,i}-\sigma^{\rm obs}_{0,i}}{\Delta \sigma_{0,i}^{\mathrm{tot}}}\right)^{2}\right]\;.
\end{equation}

\begin{figure}
\centerline{\includegraphics[keepaspectratio,clip,width=0.5\textwidth]{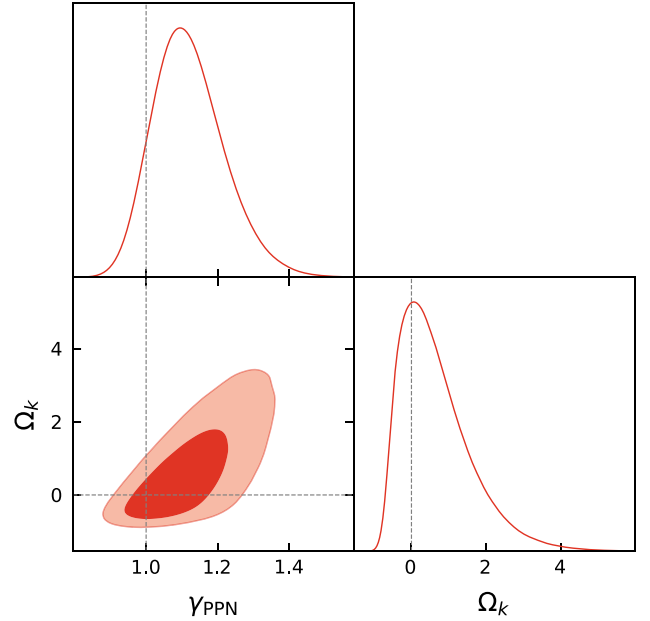}}
\vskip-0.2in
\caption{1D and 2D marginalized probability distributions with $1\sigma$ and $2\sigma$
confidence contours for the PPN parameter $\gamma_{\rm PPN}$ and cosmic curvature $\Omega_k$.
The dashed lines correspond to a flat universe with the validity of GR ($\Omega_k=0$, $\gamma_{\rm PPN}=1$).}
\label{f1}
\end{figure}

\section{Cosmology-independent Constraints on $\gamma_{\rm PPN}$ and $\Omega_k$}
\label{sec:constraint}

We obtain cosmology-independent constraints on $\gamma_{\rm PPN}$ and $\Omega_k$ by fitting the strong-lensing and SN
data simultaneously using the Python Markov Chain Monte Carlo module EMCEE \citep{2013PASP..125..306F}. The final
log-likelihood sampled by EMCEE is a sum of the likelihoods of the SGL systems and SNe Ia:
\begin{equation}
\ln\left(\mathcal{L}_{\rm tot}\right) = \ln\left(\mathcal{L}_{\rm SGL}\right)
   + \ln\left(\mathcal{L}_{\rm SN}\right)\;.
\end{equation}
The third-order polynomial modeling the distance function $d(z)$ has two free parameters ($a_1$ and $a_2$).
The absolute magnitude $M_{B}$ enters into the SN likelihood as a nuisance parameter. The PPN parameter $\gamma_{\rm PPN}$
and the lens model parameters ($\alpha_{0}$, $\alpha_{z}$, and $\alpha_{s}$) enter into the SGL likelihood as
four free parameters. In addition, the $d_{ls}/d_s$ given by Equation~(\ref{eq:DSR}) involves the curvature parameter
$\Omega_{k}$, making it eight free parameters in total.

By marginalizing the lens model parameters ($\alpha_{0}$, $\alpha_{z}$, and $\alpha_{s}$), the polynomial coefficients
($a_1$ and $a_2$), and the SN absolute magnitude $M_{B}$, we obtain the 1D and 2D marginalized probability distributions
with $1\sigma-2\sigma$ confidence regions for $\gamma_{\rm PPN}$ and $\Omega_{k}$, which are presented in Figure~\ref{f1}.
These contours show that, whereas $\Omega_{k}=0.48^{+1.09}_{-0.71}$ is weakly constrained, we can set a good limit of
$\gamma_{\rm PPN}=1.11^{+0.11}_{-0.09}$ at the 68\% confidence level. The inferred value of the PPN parameter is
compatible with the prediction of $\gamma_{\rm PPN}=1$ from GR. The constraint accuracy of $\gamma_{\rm PPN}$ is about 9.0\%.
As shown in Table~\ref{table1}, the lens model parameters are constrained to be $\alpha_{0}=1.266^{+0.105}_{-0.105}$,
$\alpha_{z}=-0.332^{+0.169}_{-0.188}$, and $\alpha_{s}=0.656^{+0.065}_{-0.065}$ at the 68\% confidence level, which are
consistent with the results of \cite{2019MNRAS.488.3745C}. We find that $\alpha_{z}=0$ is ruled out at $\sim2\sigma$ level
and $\alpha_{s}=0$ is ruled out at $\sim10\sigma$ level, confirming the significant dependencies of the total mass density
slope $\alpha$ on both the lens redshift and the surface mass density.

\begin{figure}
\centerline{\includegraphics[keepaspectratio,clip,width=0.35\textwidth]{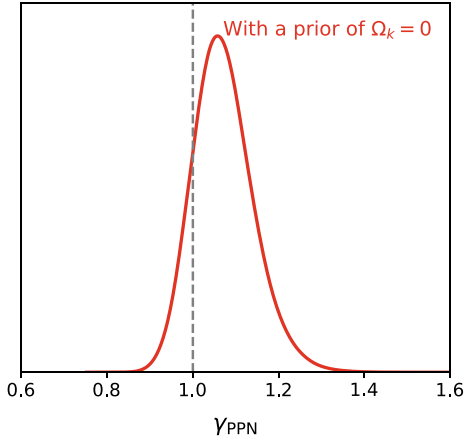}}
\vskip-0.2in
\caption{1D marginalized probability distribution of the PPN parameter $\gamma_{\rm PPN}$, assuming a flat universe.
The vertical dashed line represents the prediction of $1$ from GR.}
\label{f2}
\end{figure}

\begin{figure}
\centerline{\includegraphics[keepaspectratio,clip,width=0.35\textwidth]{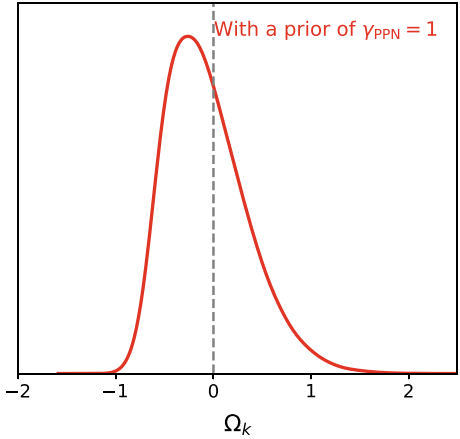}}
\vskip-0.2in
\caption{1D marginalized probability distribution of the curvature parameter $\Omega_k$, assuming GR holds on.
The vertical dashed line corresponds to a spatially flat universe.}
\label{f3}
\end{figure}

If a prior of flatness (i.e., $\Omega_k=0$) is adopted, the resulting posterior probability distribution for $\gamma_{\rm PPN}$
is shown in Figure~\ref{f2}. The result $\gamma_{\rm PPN}=1.07^{+0.07}_{-0.07}$ ($1\sigma$ confidence level) is
in good agreement with $\gamma_{\rm PPN}=1$ predicted by GR, and its constraint accuracy is improved to about 6.5\%.
If we instead assume GR holds (i.e., $\gamma_{\rm PPN}=1$), and allow $\Omega_k$ to be a free parameter, we obtain
the marginalized probability distribution for $\Omega_k$, as illustrated in Figure~\ref{f3}. The curvature parameter is
constrained to be $\Omega_k=-0.12^{+0.48}_{-0.36}$, consistent with a flat universe. The corresponding results for
all parameters are summarized in lines 1--3 of Table~\ref{table1} for the cases with no priors, the prior of $\Omega_k=0$,
and the prior of $\gamma_{\rm PPN}=1$, respectively. The comparison among these three cases indicates that the nuisance
parameters ($\alpha_{0}$, $\alpha_{z}$, $\alpha_{s}$, $a_1$, $a_2$, and $M_{B}$) have little effect on the PPN parameter
$\gamma_{\rm PPN}$ and cosmic curvature $\Omega_k$.

\begin{table*}
\centering \caption{Cosmology-independent Constraints on All Parameters from the Pantheon SN Ia and SGL Observations Using
Various Choices of Priors}
\begin{tabular}{lcccccccc}
\hline
\hline
Priors & $\gamma_{\rm PPN}$  &  $\Omega_{k}$  &  $\alpha_{0}$  &  $\alpha_{z}$   &  $\alpha_{s}$  &  $a_1$  &  $a_2$  & $M_B$ \\
\hline
None & $1.11^{+0.11}_{-0.09}$ & $0.48^{+1.09}_{-0.71}$ & $1.266^{+0.105}_{-0.105}$ & $-0.332^{+0.169}_{-0.188}$  & $0.656^{+0.065}_{-0.065}$ & $-0.245^{+0.021}_{-0.021}$ & $0.018^{+0.016}_{-0.016}$ & $-19.348^{+0.011}_{-0.011}$ \\
$\Omega_k=0$ & $1.07^{+0.07}_{-0.07}$ & -- & $1.259^{+0.103}_{-0.103}$ & $-0.238^{+0.093}_{-0.095}$  & $0.649^{+0.064}_{-0.064}$ & $-0.245^{+0.021}_{-0.021}$ & $0.017^{+0.016}_{-0.016}$ & $-19.348^{+0.011}_{-0.011}$ \\
$\gamma_{\rm PPN}=1$ & -- & $-0.12^{+0.48}_{-0.36}$ & $1.200^{+0.087}_{-0.088}$ & $-0.188^{+0.114}_{-0.120}$  & $0.674^{+0.062}_{-0.062}$ & $-0.242^{+0.021}_{-0.021}$ & $0.015^{+0.016}_{-0.016}$ & $-19.349^{+0.011}_{-0.011}$ \\
\hline
\end{tabular}
\label{table1}
\end{table*}

\section{Summary and Discussions}
\label{sec:summary}

Galaxy-scale strong-lensing systems with measured stellar velocity dispersions provide an excellent extragalactic test
of GR by constraining the PPN parameter ($\gamma_{\rm PPN}$). Measuring $\gamma_{\rm PPN}$ in this manner, however,
one has to know the lensing distances (the angular diameter distances between the source, lens, and observer),
which are conventionally calculated within the standard $\Lambda$CDM cosmological model. Because $\Lambda$CDM
itself is built on the theoretical framework of GR, these distance calculations would involve a circularity problem.
In this work, aiming to overcome the circularity problem, we have applied the DSR in the FLRW metric to obtain
cosmology-independent constraints on both $\gamma_{\rm PPN}$ and the cosmic curvature parameter $\Omega_k$.
Though the DSR method has been used to directly infer the value of $\Omega_k$ by confronting observations of
SGL systems with SN Ia luminosity distances, the simultaneous measurement of $\Omega_k$ and $\gamma_{\rm PPN}$
has not yet been achieved by the community in the literature.

Combining 120 well-measured SGL systems at $z_{s}<2.3$ with the latest Pantheon SN Ia observations, we have
simultaneously placed limits on $\gamma_{\rm PPN}$ and $\Omega_k$ without any assumptions about the contents
of the universe or the theory of gravity. This analysis suggests that the PPN parameter is constrained to be
$\gamma_{\rm PPN}=1.11^{+0.11}_{-0.09}$, representing a precision of 9.0\%, consistent with the prediction of
1 from GR at a 68\% confidence level. Meanwhile, the optimized curvature parameter is $\Omega_{k}=0.48^{+1.09}_{-0.71}$.
If using the spatial flatness as a prior, we find $\gamma_{\rm PPN}=1.07^{+0.07}_{-0.07}$, representing an agreement with
GR to 6.5\%. Assuming GR is valid and allowing $\Omega_k$ to be a free parameter, we infer that
$\Omega_k=-0.12^{+0.48}_{-0.36}$. This cosmic curvature value does not significantly deviate from
a flat universe.

Previously, \cite{2017ApJ...835...92C} obtained a 25\% precision on the determination of $\gamma_{\rm PPN}$
by analyzing a sample of 80 lenses in the flat $\Lambda$CDM model. Under the assumption of the fiducial
$\Lambda$CDM cosmology with parameters taken from Planck observations, \cite{2018Sci...360.1342C} estimated
$\gamma_{\rm PPN}$ on scales around 2 kpc to be $0.97\pm0.09$ (representing a 9.3\% precision measurement)
by using a nearby SGL system, ESO 325-G004. \cite{2020MNRAS.497L..56Y} derived $\gamma_{\rm PPN}=0.87^{+0.19}_{-0.17}$
(representing a precision of 21\%) for flat $\Lambda$CDM using a sample of four time-delay lenses.
Within the framework of the flat FLRW metric, \cite{2021arXiv210902291L} used 120 strong-lensing
data to obtain a model-independent constraint of $\gamma_{\rm PPN}=1.065^{+0.064}_{-0.074}$ (representing a
precision of 6.5\%) by implementing Gaussian processes to extract the SN distances. Despite not assuming
a specific cosmological model, the uncertainties in our constraints are comparable to these previous results.
Most importantly, our method offers a new cosmology-independent way of simultaneously constraining both
$\gamma_{\rm PPN}$ and $\Omega_{k}$.

Forthcoming lens surveys such as the Large Synoptic Survey Telescope, with improved depth, area, and resolution,
will be able to increase the current galactic-scale lens sample sizes by orders of magnitude \citep{2015ApJ...811...20C}.
With such abundant observational information in the future, the mass-dynamical structure of the lensing galaxies
will be better characterized, and model-independent constraints on the PPN parameter $\gamma_{\rm PPN}$ and
cosmic curvature $\Omega_{k}$, as discussed in this work, will be considerably improved.

Finally, we investigated whether the approximation of the dimensionless distance function $d(z)$
(as a linear polynomial; see Equation~\ref{eq:dz}) affects the inference of $\gamma_{\rm PPN}$.
To probe the dependence of the outcome on the approximation of $d(z)$, we also performed a parallel comparative
analysis of the SGL and SN Ia data using the exact expression in the flat $\Lambda$CDM model, i.e.,
$d(z)=\int_{0}^{z}\frac{\mathrm{d}z'}{\sqrt{\Omega_{\rm m}(1+z')^{3}+1-\Omega_{\rm m}}}$. In this case,
the free parameters are the PPN parameter $\gamma_{\rm PPN}$, the lens model parameters ($\alpha_{0}$, $\alpha_{z}$,
and $\alpha_{s}$), the matter density parameter $\Omega_{\rm m}$, and the SN absolute magnitude $M_{B}$.
We found that the constraints are $\gamma_{\rm PPN}=1.07^{+0.07}_{-0.07}$, $\alpha_{0}=1.254^{+0.103}_{-0.103}$,
$\alpha_{z}=-0.232^{+0.090}_{-0.093}$, $\alpha_{s}=0.653^{+0.064}_{-0.064}$, $\Omega_{\rm m}=0.302^{+0.022}_{-0.022}$,
and $M_{B}=-19.350^{+0.011}_{-0.011}$. Comparing these inferred parameters with those obtained with the linear
polynomial fit (see line 2 in Table~\ref{table1}), it is clear that the linear polynomial function provides
a good approximation of $d(z)$ and the adoption of the exact expression for $d(z)$ in the flat $\Lambda$CDM model
only has a minimal influence on these results.

\acknowledgments
We would like to thank the anonymous referee for helpful comments.
This work is partially supported by the National Natural Science Foundation of China (grant Nos.~11988101, 11725314, U1831122, 12041306, 11633001, 11920101003, 12021003, and 12033008), the Youth Innovation Promotion Association (2017366), the Key Research Program of Frontier Sciences (grant No. ZDBS-LY-7014) of Chinese Academy of Sciences, the Strategic Priority Research Program of the Chinese Academy of Sciences (grant No. XDB23000000), the Major Science and Technology Project of Qinghai Province (2019-ZJ-A10), the China Manned Space Project (Nos. CMS-CSST-2021-B11, CMS-CSST-2021-B01, and CMS-CSST-2021-A01),
the Guangxi Key Laboratory for Relativistic Astrophysics, the K. C. Wong Education Foundation, and the Interdiscipline Research Funds of Beijing Normal University.


\end{document}